\documentclass[superscriptaddress,secnumarabic,amssymb,amsmath,nobibnotes,aps,prd,showkeys,showpacs,nofootinbib,twocolumn]{revtex4}%
\usepackage{graphicx}
\usepackage{epsf}
\usepackage{bm}
\usepackage{amsmath}
\usepackage{amsfonts}
\usepackage{amssymb}
\usepackage{epstopdf}
\usepackage{natbib}%
\usepackage{color}

\begin{document}

\title{A note on the relations between thermodynamics, energy definitions and Friedmann equations}

\author{Hooman Moradpour}
\email{h.moradpour@riaam.ac.ir}
\affiliation{Research Institute for Astronomy and Astrophysics of Maragha (RIAAM), P.O. Box 55134-441, Maragha, Iran}
\author{Rafael C. Nunes}
\email{rafadcnunes@gmail.com;nunes@ecm.ub.edu}
\affiliation{Departamento de F\'isica, Universidade Federal de Juiz de Fora, 36036-330, Juiz de Fora, MG, Brazil}
\author{Everton M. C. Abreu}
\email{evertonabreu@ufrrj.br}
\affiliation{Departamento de F\'isica, Universidade Federal de Juiz de Fora, 36036-330, Juiz de Fora, MG, Brazil}
\affiliation{Grupo de Física Teórica e Matemática Física, Departamento de Física, Universidade Federal Rural do
Rio de Janeiro, 23890-971 Seropédica, Rio de Janeiro, Brazil}
\author{Jorge Ananias Neto}
\email{jorge@fisica.ufjf.br}
\affiliation{Departamento de F\'isica, Universidade Federal de Juiz de Fora, 36036-330, Juiz de Fora, MG, Brazil}

\pacs{ }

\begin{abstract}
\noindent In what follows, we investigate the relation between the
Friedmann and thermodynamic pressure equations, through solving
the Friedmann and thermodynamic pressure equations simultaneously.
Our investigation shows that a perfect fluid, as a suitable
solution for the Friedmann equations leading to the standard
modeling of the universe expansion history, cannot simultaneously
satisfy the thermodynamic pressure equation and those of
Friedmann. Moreover, we consider various energy definitions, such
as the Komar mass, and solve the Friedmann and thermodynamic
pressure equations simultaneously to get some models for dark
energy. The cosmological consequences of obtained solutions are
also addressed. Our results indicate that some of obtained
solutions may unify the dominated fluid in both the primary
inflationary and current accelerating eras into one model. In
addition, by taking into account a cosmic fluid of a known
equation of state, and combining it with the Friedmann and
thermodynamic pressure equations, we obtain the corresponding
energy of these cosmic fluids and face their limitations. Finally,
we point out the cosmological features of this cosmic fluid and
also study its observational constraints.
\end{abstract}

\maketitle

\section{Introduction}\label{intro}

The homogeneity and isotropy of cosmos in scales larger than
$100$-Mpc motivates us to use FLRW metric and an energy-momentum
source in the $T^{\mu}_{\nu}=diag(\rho,p,p,p)$ form to describe
the spacetime geometry, where $\rho$ and $p$ are the energy
density and momentum of the source, respectively \cite{roos}.
$p=w\rho$ is a simple solution, used to describe the universe
expansion eras, in which $w$ is called the state parameter, and
$w=\frac{1}{3},0$ denote the radiation and matter dominated eras,
respectively \cite{roos}. Moreover, a fluid with
$w\leq-\frac{2}{3}$ and $w=-1$ is needed in order to describe the
current and primary inflationary phases of the universe,
respectively \cite{roos,mr}. It has also been shown that a perfect
fluid model of current accelerating phase cannot satisfy the
thermodynamics stability conditions \cite{rafael}. Here, we should
note that, although a perfect fluid with $w\leq-\frac{2}{3}$
cannot satisfy the thermodynamic stability conditions
\cite{mr,rafael}, the whole system, including the geometry and the
cosmic fluid filling the background, meets the stability
conditions \cite{mr,msrw,pavr,pavr1}. The latter is due to the
effects of the cosmic dominated fluid on the horizon part of total
entropy which controls the behavior of total entropy in the long
run limit (the current phase of universe)
\cite{mr,msrw,pavr,pavr1}. In fact, the universe meets the
thermodynamic equilibrium, whenever the dominated cosmic fluid is
a prefect fluid of $-1\leq w\leq-\frac{2}{3}$
\cite{mr,msrw,pavr,pavr1}.

In addition, to describe the universe' dynamics, some authors use
a type of fluid with the equation of state $p=p(\rho)$, which
differs from the perfect fluid model
\cite{od1,nojp,od2,od3,poly1,ref57,nK,poly2,poly20,poly200,chap1,chap10,chap2,chap3,van,jenk,jenk1,nucl,quark,prd0,capa,wang}.
Such equation of state can be generated from the quantum effects
near the future singularities \cite{od1,nojp,od2,od3} and also the
so-called bag models of the hadron production from the
quarks-gluons coagulation \cite{jenk,jenk1,nucl,quark}. In fact, a
fluid of varying state parameter may lead to a unified description
for the cosmos sectors such as dark matter and dark energy
\cite{prd0,capa}. It may also help us in finding out a better
description for the universe transitions between its various eras
\cite{prd0,capa}. Moreover, considering such equation of state,
one can get a suitable description for the interactions between
the dark energy and other parts of cosmos which may solve the
coincidence problem \cite{capa,wang}. Therefore, it is of the
great importance to study the cosmological features of fluids of
non-constant state parameter \cite{prd0,capa,wang}. It is also
shown that fluids with two free parameters, obeying the van der
Waals equation of state or $p=k\rho^{\gamma}$, where $k$ and
$\gamma$ are constants, may satisfy the thermodynamics stability
conditions \cite{plb1,plb2,ptp,abri}. Since the FLRW universe is a
spherically symmetric spacetime, it seems that the Misner-Sharp
mass is a proper definition for the total energy confined by the
FLRW apparent horizon
\cite{ms,Hay22,Hay2,Bak,caiwork,sheyw1,sheyw2,ufl,md}. Moreover,
one can also combine the Komar mass definition with the first and
second laws of thermodynamics as well as the holographic principle
to obtain the Einstein and thus Friedmann equations and their
various modifications \cite{ver,smr,msh,pad,cair,me}. In fact,
there are many other mass definitions used to investigate cosmic
fluids \cite{roos,moremass}.

Friedmann equations relate the energy density and pressure of
cosmic fluid to the spacetime parameters. These equations do not
say a thing about the nature of the source and in fact, they can
only tell how the source parameters, including its energy density
and pressure, should behave. Therefore, one may assume a special
fluid, such as the perfect fluid, and solve the equations to
obtain the source parameters as the functions of the spacetime
parameters, such as the Hubble parameter \cite{roos}. In addition,
it seems that the apparent horizon of the FLRW metric plays the
role of causal boundary for this spacetime, and thus may be
considered as a proper boundary to investigate the thermodynamics
of system \cite{Hay22,Hay2,Bak}. As a matter of fact, one can show
that the unified first law of thermodynamics is valid on this
surface, a result which is parallel to the validity of the
Friedmann equations
\cite{Hay22,Hay2,Bak,caiwork,sheyw1,sheyw2,ufl}.

Now, taking into account this horizon as a boundary for the
system, it can be shown that, as we have previously mentioned, the
thermodynamic stability conditions of closed systems
\cite{CALEN,path}, including the universe and the fields that fill
it, are met \cite{mr,msrw,pavr,pavr1}. In fact, since the
observational data permit a dominated fluid of $-1\leq
w\leq-\frac{2}{3}$ in the current stage of the universe
\cite{roos}, the thermodynamic stability conditions are preserved
by the universe \cite{mr,msrw,pavr,pavr1} meaning that the current
state of the universe may be a thermodynamic equilibrium state.
Thus, in the current state of the universe, the expectation of the
availability of the thermodynamic intensive quantities such as the
pressure and temperature for the dominated fluid is not unlikely.
Therefore, if we look at the current state (the long run limit) of
the cosmos as a thermodynamical system in equilibrium
\cite{mr,msrw,pavr,pavr1}, at least in the time scale of human
life, then we may also consider a thermodynamic equilibrium
pressure for the cosmic fluid. The latter means that a solution
for the Friedmann equations may also satisfy the thermodynamic
pressure equation which is indeed a thermodynamic equation of
state \cite{CALEN,path}. Because the thermodynamic pressure is
defined as the derivative of energy with respect to the system
volume \cite{CALEN,path}, we need the cosmic fluid energy relation
to investigate the relation between the thermodynamic pressure and
Friedmann equations. We finally think that searching for such
solutions, satisfying the thermodynamic pressure and Friedmann
equations simultaneously, may also help us to come close to a more
proper definition for the energy of cosmic fluid.

Our aim in this paper is to investigate the relation between the
Friedmann equations, the thermodynamic pressure and various energy
definitions. In fact, we are going to find some solutions for the
Friedmann equations, which at the same time would satisfy the
thermodynamic pressure definition, and use them to model the dark
energy candidates. In order to do so, taking the various energy
definitions of cosmic fluid, on one hand, we use the thermodynamic
pressure equation to obtain the corresponding pressure and then
compare the result to those obtained from solving the Friedmann
equations. Moreover, by considering a general form for the energy,
and combining it with the thermodynamic pressure equation as well
as the Friedmann equations, we find some new solutions. On the
other hand, focusing on the situation in which the equation of
state is known, we will try to find those new solutions for the
Friedmann equations through which both the Friedmann equations and
the thermodynamic pressure equation can be satisfied
simultaneously. This may help us to find the corresponding energy
of this solution in the FLRW background. We also take advantage of
some observation data to study the cosmological consequences of
the solutions we obtained.

We organize the paper as follows. In the next section, bearing the
various energy definitions in mind, we show that a perfect fluid
model cannot satisfy both the Friedmann and thermodynamic pressure
equations simultaneously. This motivates us to use other fluids
and energy definitions in our modeling of the cosmic fluid. In
fact, this section helps us in clarifying our purpose and
motivation in the presentation of this work. In the section
$\textmd{III}$, we will consider some energy definitions and
combine them with the thermodynamic pressure and Friedmann
equations to have consistent solutions for the cosmic fluid, and
use the obtained solutions for modeling dark energy. In fact, our
recipe, in this section, may help us to provide a thermodynamic
motivation for some type of cosmic fluids and thus the dark energy
candidates. In section $\textmd{IV}$, we will focus on the fluids
with a known equation of state, and we will try to solve the
Friedmann and thermodynamic pressure equations simultaneously in
order to obtain some properties of the energy-momentum source such
as its energy and its pressure. We also point out the
observational constraints on this kind of fluids for modeling the
dark energy in section $\textmd{V}$. Throughout the paper, the
cosmological consequences of the obtained solutions for describing
the current phase of the universe will be also addressed. Section
$\textmd{VI}$ is dedicated to a summary of the results and
concluding remarks.

\section{perfect fluid model cannot satisfy thermodynamics and cosmology simultaneously}

The metric of the FLRW universe with scale factor $a(t)$ is
written as

\begin{eqnarray}\label{frw}
ds^{2}=-dt^{2}+a^{2}\left( t\right) \left[ \frac{dr^{2}}{1-\kappa r^{2}}%
+r^{2}d\Omega^{2}\right],
\end{eqnarray}

\noindent in which $\kappa=-1,0,1$, called the curvature constant,
points to the open, flat and closed universes, respectively
\cite{roos}. Apparent horizon of this spacetime, a proper causal
boundary for thermodynamic investigations, as the marginally
trapped surface is evaluated by

\begin{eqnarray}\label{ah2}
\partial_{\mu}\tilde{r}\partial^{\mu}\tilde{r}=0\rightarrow r_A,
\end{eqnarray}
where $\tilde{r}=a(t)r$, which finally leads to
\begin{eqnarray}\label{ah}
\tilde{r}_A=a(t)r_A=\frac{1}{\sqrt{H^2+\frac{\kappa}{a(t)^2}}},
\end{eqnarray}
for the physical radii of apparent horizon ($\tilde{r}_A$)
\cite{Hay2,Hay22,Bak,sheyw1,sheyw2}. Since WMAP data confirms a
flat universe, we focus on the $\kappa=0$ case, and therefore,
$V=\frac{4\pi}{3H^3}$ is the volume of the flat FLRW universe
confined by the apparent horizon located at
$\tilde{r}_A=\frac{1}{H}$. The Friedmann first equation and the
continuity equation are
\begin{eqnarray}\label{fried1}
H^2=\frac{8\pi}{3}\rho,
\end{eqnarray}
and
\begin{eqnarray}\label{cont1}
\dot{\rho}+3H(\rho+p)=0,
\end{eqnarray}
respectively. In these equations, $\rho$ and $p$ denote the energy
density and pressure of a homogenous and isotropic source
supporting the FLRW geometry, respectively. If we use
$V=\frac{4\pi}{3H^3}$ to rewrite Eq.~(\ref{fried1}), we will have
\begin{eqnarray}\label{rhov}
\rho(V)=\frac{1}{2}\Big(\frac{3}{4\pi}\Big)^{\frac{1}{3}}V^{-\frac{2}{3}},
\end{eqnarray}
in the flat FLRW universe. One can combine Eqs.~(\ref{fried1})
and~(\ref{cont1}) with each other to obtain the Friedmann second
equation as
\begin{eqnarray}\label{fried2}
3H^2+2\dot{H}=-8\pi p.
\end{eqnarray}

Therefore, a solution for $\rho$ and $p$ which satisfies
Eqs.~(\ref{fried1}) and~(\ref{cont1}) simultaneously, also meets
Eq.~(\ref{fried2}). These equations can also be combined with each
other to yield the Raychaudhuri equation
\begin{eqnarray}\label{ray}
\dot{H}=-4\pi(\rho+p).
\end{eqnarray}

A perfect fluid with $p=w\rho$, where $w$ is called the state
parameter, a constant parameter, is a simple solution for the
above equations in a good agreement with observations data
\cite{roos}. Moreover, for metric~(\ref{frw}), since it is a
spherically symmetric metric, the Misner-Sharp mass \cite{ms}
leads to $E_{MS}=\frac{1}{2H}=\rho V$ and it is used in various
papers in order to obtain the Friedmann equations from
thermodynamics arguments \cite{md,caiwork,ufl}. In fact, the
$E_{MS}=\rho V$ equality is only valid in the flat FLRW universe,
governed by the Friedmann equations, which means that
Eq.~(\ref{rhov}) is valid. As we have previously mentioned, the
dark energy candidate, as the dominated fluid in the present state
of the universe, is the backbone of the probable thermodynamic
equilibrium of the universe \cite{mr,msrw,pavr,pavr1}. In fact,
since the satisfaction of the thermodynamic equilibrium conditions
in the current stage of the universe expansion is not impossible
\cite{mr,msrw,pavr,pavr1,plb1,plb2,ptp,abri}, it is not also
unlikely to define the thermodynamic pressure for at least the
dark energy component as the dominated fluid in the current state
of the universe. In thermodynamics, pressure is defined as
\begin{eqnarray}\label{pres1}
p=-\Big(\frac{\partial E}{\partial V}\Big)_S,
\end{eqnarray}
where $E$ and $S$ are the energy and entropy of energy-momentum
source, respectively \cite{CALEN,path}. Now, for the flat universe
with $V=\frac{4\pi}{3H^3}$, by using Eq.~(\ref{fried1}) and the
Misner-Sharp mass to compute pressure from Eq.~(\ref{pres1}), we
have
\begin{eqnarray}\label{pres2}
p=-\frac{1}{3}\rho,
\end{eqnarray}
independently of the nature of the energy-momentum source. It goes
without saying that this result is in direct conflict with the
universe expansion history, for example, the radiation dominated
era. In short, this result indicates that the sign of the
thermodynamic pressure differs from the one of the energy density,
a result which is in agreement with previous works showing that
the thermodynamic pressure of a universe filled by an ideal gas
with vanishing speed of sound can only be zero or negative
\cite{luo}. Now, since some authors have shown that the Komar mass
($E=(\rho+3p)V$) is in agreement with the Friedmann and continuity
equations as well as the thermodynamics laws
\cite{ver,smr,pad,cair,msh}, we may consider it as a true mass
definition in the cosmological setups. By following the above
recipe for the Komar mass and a perfect fluid with constant $w$,
we see that only a perfect fluid with $w=-\frac{1}{6}$ meets all
of the above equations, which seems unsatisfactory. Moreover, if
we use the $(\rho+p)V$ definition of energy \cite{roos}, we obtain
a perfect fluid with $w=-\frac{1}{4}$ which is again in conflict
with the universe history. Now, if we define $E=(\beta\rho+\zeta
p)V$ and follow the above recipe for a perfect fluid with constant
state parameter, we can reach the $w=-\frac{\beta}{\zeta+3}$
relation for the state parameter. In addition, since $p=w\rho$, we
have $E=(\beta+\zeta w)\rho V$ which finally leads to $E=-3 w \rho
V$. As it is obvious, a perfect fluid with a positive constant
state parameter, such as radiation, cannot lead to solutions with
positive energy in the universe whole history, a result which
again looks unsatisfactory. Indeed, the latter result tells us
that since $E$ is a positive quantity, $w$ should meet the $w \leq
0$ condition, a result which may be in line with the current
accelerating universe \cite{roos}, but it is in direct conflict
with the radiation and matter dominated eras as well as the
thermodynamics stability conditions \cite{rafael}. Therefore,
although a perfect fluid with $p= w \rho$ helps us in describing
the universe expansion in an appropriate manner, it cannot satisfy
the Friedmann, continuity and thermodynamic pressure equations as
well as the $E>0$ condition simultaneously. Indeed, one may
conclude that there is an intrinsic inconsistency in the triple
relation between thermodynamics, Friedmann equations and thus the
universe history, if one takes into account the perfect fluid
concept, as the dominant perfect fluid, and the mentioned energy
definitions for modeling the universe dynamics. This inconsistency
motivates us to find and use another energy definitions and cosmic
fluids which can always meet the Friedmann, continuity and the
thermodynamic pressure equations simultaneously.

\section{From energy definition, Friedmann and thermodynamic equations to the cosmic fluid}

Here, we combine the energy definitions with the Friedmann and
thermodynamic pressure equations to construct a model for the
cosmic fluid.

\subsection{The Komar mass}

Inserting the Komar mass, $E=(\rho+3p)V$, and Eq.~(\ref{rhov})
into Eq.~(\ref{pres1}) and using $\frac{\partial}{\partial
V}=\frac{\partial}{\partial\rho}\frac{\partial\rho}{\partial V}$
we have that

\begin{eqnarray}\label{pres03}
\frac{dp}{d\rho}-\frac{2p}{\rho}-\frac{1}{6}=0,
\end{eqnarray}
which leads to

\begin{eqnarray}\label{pres3}
p(\rho)=\rho \Big(c_0\rho-\frac{1}{6} \Big),
\end{eqnarray}
where $c_0$ is an integration constant. This result may cover a
modified Polytropic model with the Polytropic index $n=1$
\cite{poly1}, and also a modified Chaplygin model with the
Chaplygin parameter $\alpha=-2$ \cite{chap1,chap10}, which are
used to describe the nature of dominant fluid in current phase of
the universe expansion. Here, it is also interesting to note that,
as a check, the result of a perfect fluid with $w=-\frac{1}{6}$,
obtained in previous section, is produced in the $c_0\rightarrow0$
limit. Nevertheless,, inserting Eq.~(\ref{pres3}) into
Eq.~(\ref{cont1}) we find

\begin{eqnarray}\label{den1}
\rho_x(a)=\rho_{x0}\frac{5a^{\frac{-5}{2}}}{1-6c_0a^{\frac{-5}{2}}}.
\end{eqnarray}


Therefore, $E=(\rho+3p)V$ is the energy content of a universe
where the energy-momentum source satisfies Eqs.~(\ref{pres3})
and~(\ref{den1}). We note here that the authors in
\cite{poly1,chap1,chap10}, use the Misner-Sharp definition to
study their models, while we have seen that the Komar mass is a
more suitable definition of energy for these models. Finally, our
work proposes that,whenever Komar mass is combined with the
thermodynamics, Friedmann and continuity equations, it will
provide a motivation for modeling the flat FLRW universe by a
source satisfying Eq.~(\ref{pres3}), which is similar to a
modified Polytropic gas with $n=1$ and a modified Chaplygin model
with $\alpha=-2$.
\\


Let us now investigate the cosmological consequences of the
solution given by Eq.~(\ref{den1}) that, associated with the
thermodynamic pressure Eq.~(\ref{pres3}) results in an equation of
state (EoS) defined by

\begin{eqnarray}\label{w1}
w(z)= \frac{p}{\rho} = \bar{c_0}
\frac{5(1+z)^{\frac{5}{2}}}{1-6c_0(1+z)^{\frac{5}{2}}} -
\frac{1}{6},
\end{eqnarray}
where we have defined $\bar{c_0} = c_0 \rho_{x0}$. From
Eq.~(\ref{den1}), one can check that $\rho_x(a)$ tends to the
constant $\rho_{x0}$ for the current value of scale factor
($a\rightarrow1$), while $c_0=-\frac{4}{6}$. Fig.~($1$) shows the
evolution of $w(z)$ for several values of $\bar{c_0}$.
\\
\begin{figure}[ht]
\centering
\includegraphics[scale=0.4]{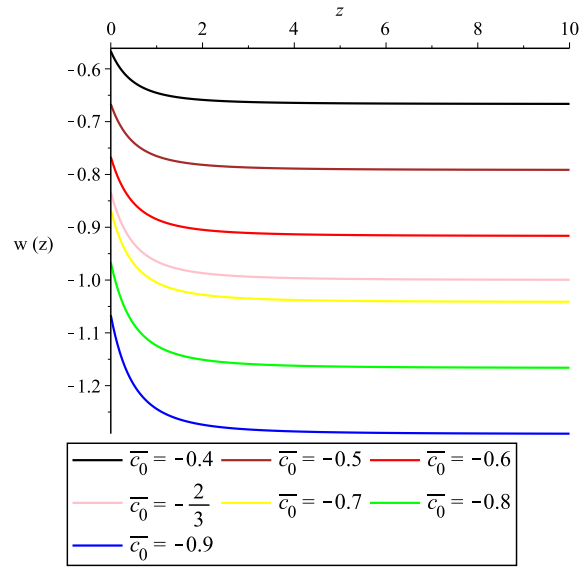}
\caption{EoS given by Eq. (\ref{w1}) as a function of the redshift
for some values of $\bar{c_0}$.}
\end{figure}
It is worth mentioning that for $z\geq0$, including both the past
and present state of universe, $w(z)$ is always negative for
$c_0=-\frac{4}{6}$. Therefore, one may use $w(z)$ to describe the
primary inflationary and current accelerating expansions of
universe. As an example, for
$-\frac{5}{6}\leq\bar{c_0}\leq-\frac{1}{2}$, EoS is within the
$-1\leq\omega(z)\leq-\frac{2}{3}$ range in the $z\rightarrow0$
limit. Since $\bar{c_0} = c_0 \rho_{x0}$, one can finally has the
$\frac{3}{4}\leq\rho_{x0}\leq\frac{5}{4}$ range for $\rho_{x0}$ in
this situation. The $\bar{c_0}=-\frac{2}{3}$ case leads to
interesting behavior. For this case, $w(z)$ tends to $-1$ and
$-\frac{5}{6}$ as $z\rightarrow\infty$ and $z\rightarrow0$,
respectively. Therefore, this case may be used to model the
dominated fluids in both the primary inflationary and current
accelerating phases into one model.




The Friedmann's equation (\ref{fried1}) in the presence of the
fluid given by Eq.~(\ref{den1}) is written as

\begin{eqnarray}\label{H1}
 \frac{H^2(z)}{H^2_0} = \Omega_{m0} (1+z)^3 + \Omega_{x0} \frac{5(1+z)^{\frac{5}{2}}}{1-6c_0(1+z)^{\frac{5}{2}}},
\end{eqnarray}
where, taking $H(z=0)=H_0$, we have the constraints $\Omega_{x0} =
1 - \Omega_{m0}$.
\\

Since the deceleration parameter defined as,
$$q=-1-\frac{dH}{da}\frac{a}{H},$$ using Eq.~(\ref{H1}) and
inserting $c_0=-\frac{4}{6}$ we find
\begin{eqnarray}\label{q00}
q(z)=\frac{\frac{5 (1 - \Omega_{m0})
[1-16(1+z)^{\frac{5}{2}}]}{\Omega_{m0}(1+z)^{\frac{1}{2}}}+2[1+4(1+z)^{\frac{5}{2}}]^2}
{\frac{20(1 -
\Omega_{m0})[1+4(1+z)^{\frac{5}{2}}]}{\Omega_{m0}(1+z)^{\frac{1}{2}}}+4[1+4(1+z)^{\frac{5}{2}}]^2},
\end{eqnarray}
which is plotted in Fig.~($2$) for $\Omega_{m0}=0.31$. It tends to
$\frac{1}{2}$ and $-0.362$ as $z\rightarrow\infty$ and
$z\rightarrow0$, respectively. Transition to negative deceleration
parameters happens around $z\simeq0.68$ which is close to
observation ($z\simeq0.6$) \cite{roos}.
\begin{figure}[ht]
\centering
\includegraphics[scale=0.4]{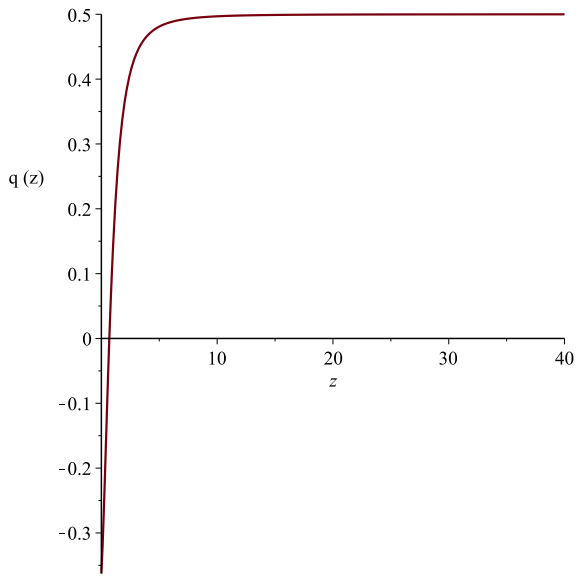}
\caption{Evolution of the deceleration parameter $q(z)$ defined in
Eq. (\ref{q00}). In drawing the graphs we have taken $\Omega_{m0}
= 0.31$.}
\end{figure}

The universe can admit a future singularity, if the cosmic fluid
satisfies at least one of the below conditions \cite{nojp,od3}:

\begin{description}

\item[Type $\textmd{I}$ singularity:] $\rho,|p|\rightarrow\infty$
for $t\rightarrow t_c$ and $a\rightarrow\infty$, \item[Type
$\textmd{II}$ singularity:] $\rho\rightarrow\rho_c$ and
$|p|\rightarrow\infty$, while $t\rightarrow t_c$ and $a\rightarrow
a_c$. \item[Type $\textmd{III}$ singularity:]
$\rho,|p|\rightarrow\infty$, for $t\rightarrow t_c$ and
$a\rightarrow a_c$, \item[Type $\textmd{IV}$ singularity:]
$\rho,|p|\rightarrow0$, for $t\rightarrow t_c$ and $a\rightarrow
a_c$,
\end{description}
where $t_c$, $\rho_c$ and $a_c\neq0$ are constant. A phantom
source leads to the type $\textmd{I}$ singularity called Big Rip
\cite{bigrip}. The second type, called sudden,  happens while $p$
diverges at finite $\rho$, $a$ and thus time \cite{sudden}. The
third type will happen whenever both $\rho$ and $p$ diverge at
finite $a$ and thus a finite time \cite{od2}. Type $\textmd{IV}$
singularity may also be reached while both $\rho$ and $p$ vanish
at finite time and thus a finite scale factor $a$ \cite{nojp}. The
last two types may be appeared in a universe supported by a source
with $p=-\rho-f(\rho)$ equation of state \cite{nojp,od2,od3}. It
is easy to check that Eq.~(\ref{den1}) avoids any future
singularity for $c_0<0$, that means the $c_0=-\frac{4}{6}$ case is
free from singularity.
\\

\subsection{The general proposal for energy and its consequences}

Now, bearing the homogeneity and isotropy of the universe in mind,
we may write the energy content of the universe as $E=\beta\rho
V+\zeta pV$, which is in fact a generalization of the mentioned
energy definitions. It covers the Misner-Sharp in the $\zeta=0$
and $\beta=1$ limits. Moreover, it converges to the Komar mass by
inserting the $1$ and $3$ values for $\beta$ and $\zeta$
parameters, respectively. Another interesting case is the
$\beta=-\zeta=\frac{1}{2}$ case leading to $E=WV$, where
$W=\frac{\rho-p}{2}$ is the work density \cite{caiwork}. The
latter case is interesting since it includes the work density
which plays the role of pressure in deriving the Friedmann
equations by applying the unified first law of thermodynamics on
the various horizons of FLRW universe
\cite{Hay22,Hay2,Bak,caiwork,sheyw1,sheyw2,ufl,md}.

Following the above recipe for $E=\beta\rho V+\zeta pV$, we obtain
that
\begin{eqnarray}\label{pressub}
\frac{dp}{d\rho}-\frac{3(\zeta+1)p}{2\zeta\rho}-\frac{\beta}{2\zeta}=0,
\end{eqnarray}
which leads to
\begin{eqnarray}\label{pressub1}
p(\rho)=A\rho^{\frac{3(\zeta+1)}{2\zeta}}-\frac{\beta\rho}{\zeta+3},
\end{eqnarray}
where $A$ is an integration constant. Now, inserting the latter
equation into~(\ref{cont1}), we reach
\begin{eqnarray}\label{pressub2}
\rho(a)=\rho_{x0}[\frac{(\zeta+3-\beta)a^{-\frac{3(\zeta+3-\beta)}{2\zeta}}}{1-A(3+\zeta)a^{-\frac{3(\zeta+3-\beta)}{2\zeta}}}]^{\frac{2\zeta}{3+\zeta}}.
\end{eqnarray}
Moreover, for the state parameter and the Friedmann equation in
the presence of such fluid, we have
\begin{eqnarray}\label{pressub3}
w(z)=\bar{A_1}[\frac{(\zeta+3-\beta)(1+z)^{\frac{3(\zeta+3-\beta)}{2\zeta}}}{1-A(3+\zeta)(1+z)^{\frac{3(\zeta+3-\beta)}{2\zeta}}}]-\frac{\beta}{\zeta+3},
\end{eqnarray}
where $\bar{A_1}=A\rho_{x0}^{\frac{\zeta+3}{2\zeta}}$, and
\begin{eqnarray}\label{pressub4}
&\frac{H^2(z)}{H^2_0}&=\Omega_{m0}(1+z)^3+\nonumber\\
&&\Omega_{x0}(\frac{(\zeta+3-\beta)(1+z)^{\frac{3(\zeta+3-\beta)}{2\zeta}}}{1-(\beta-\zeta-2)(1+z)^{\frac{3(\zeta+3-\beta)}{2\zeta}}})^{\frac{2\zeta}{3+\zeta}},
\end{eqnarray}
respectively. To obtain the last equation, we have considered
$A=\frac{\beta-\zeta-2}{3+\zeta}$ which comes from the
$\rho(z=0)=\rho_{x0}$ assumption. Calculations for the
deceleration parameter lead to
\begin{eqnarray}\label{pressub5}
q(z)=\frac{1+A(z)}{2(1+B(z))},
\end{eqnarray}
in which
\begin{eqnarray}\label{pressub6}
&&A(z) \nonumber \\
&=&\frac{\Omega_{x0}(\zeta+3-\beta)^{\frac{2\zeta}{3+\zeta}}(1+z)^{\frac{3(\zeta+3-\beta)}{3+\zeta}}}{\Omega_{m0}(1+z)^3\Big[1-(\beta-\zeta-2)(1+z)^{\frac{3(\zeta+3-\beta)}{2\zeta}}\Big]^{\frac{3(\zeta+1)}{3+\zeta}}}\nonumber \\
&\times&[2(\beta-\zeta-2)(1+z)^{\frac{3(\zeta+3-\beta)}{2\zeta}}+\frac{\zeta-3\beta+3}{3+\zeta}] \nonumber \\
\mbox{} \\
&&B(z) \nonumber \\
&=&\frac{\Omega_{x0}(\zeta+3-\beta)^{\frac{2\zeta}{3+\zeta}}(1+z)^{\frac{3(\zeta+3-\beta)}{3+\zeta}}}{\Omega_{m0}(1+z)^3\Big[1-(\beta-\zeta-2)(1+z)^{\frac{3(\zeta+3-\beta)}{2\zeta}}\Big]^{\frac{2\zeta}{3+\zeta}}}.\nonumber
\end{eqnarray}

It is easy to check that, as a desired result, the previous
results are obtainable by inserting $\beta=1$ and $\zeta=3$ into
the above equations.

%
%

This solution my cover the Polytropic and modified Polytropic
models with the Polytropic index $n=\frac{2\zeta}{3+\zeta}$ for
$\beta=0$ and $\beta\neq0$, respectively \cite{poly1}. The
Chaplygin and modified Chaplygin models with the Chaplygin
parameter $\alpha=-\frac{3(\zeta+1)}{2\zeta}$ may also be obtained
for $\beta=0$ and $\beta\neq0$, respectively \cite{chap1,chap10}.
Here, it is worth mentioning that, for $\beta=\zeta+1$, the
preceding equation of state converges to a model which was
initially introduced in~\cite{od1} and studied in more details
in~\cite{od2,nojp,od3}. More studies on the various properties of
this equation of state, such as its future singularities etc., can
be found in \cite{od2,nojp,od3,jenk,jenk1,nucl}. Therefore, our
work may give us the energy content of their model and a
thermodynamic motivation for that. It is also useful to mention
here that the quark bag model \cite{quark} permits the same
relation as Eq.~(\ref{pressub1}), which leads to the obeyance of
the results in the cosmological set-ups \cite{nucl,jenk}, and
therefore, our scheme can be considered as a thermodynamic
motivations for their claims and equations of state. In addition,
our work gives us the energy content of a universe filled by such
component. It is also interesting to mention here that the authors
in \cite{jenk} have considered a fluid with pressure
$p=\rho(a-f(\rho))$, while $-1<a$, and study its cosmological
applications. They find out that the $f(\rho)\sim\rho^{\mu}$ case,
while $\mu>0$, may lead to compatible results with the universe
history and its current phase \cite{jenk}. Let us compare our
results with those of Ref.~\cite{jenk} a little more. Comparing
their result ($\mu>0$) with Eq.~(\ref{pressub1}), we have obtained
that $\frac{\zeta+1}{\zeta}>0$ condition.

Therefore, by this scheme, we found the equation of state for the
cosmic fluid with $T^{\mu}_\nu=diag(-\rho,p,p,p)$, filling the
background, and its corresponding energy in a compatible way with
the thermodynamic pressure, Friedmann, continuity, and
Raychaudhuri equations. Finally, we should mention our
investigation shows that models proposed
in~\cite{od2,nojp,od3,jenk,jenk1,nucl} can satisfy the Friedmann
and thermodynamic equations simultaneously. In Fig.~($3$), $w(z)$
has been plotted for the some values of $\beta$, $\zeta$ and
$\bar{A_1}$ while $A=\frac{\beta-\zeta-2}{3+\zeta}$. In fact,
since we need a fluid of $\omega\leq-\frac{2}{3}$ to model the
current phase of universe \cite{roos,mr}, we should have
$\bar{A_1}-\frac{\beta}{\zeta+3}\leq-\frac{2}{3}$ at $z=0$.
\begin{figure}[ht]
\centering
\includegraphics[scale=0.4]{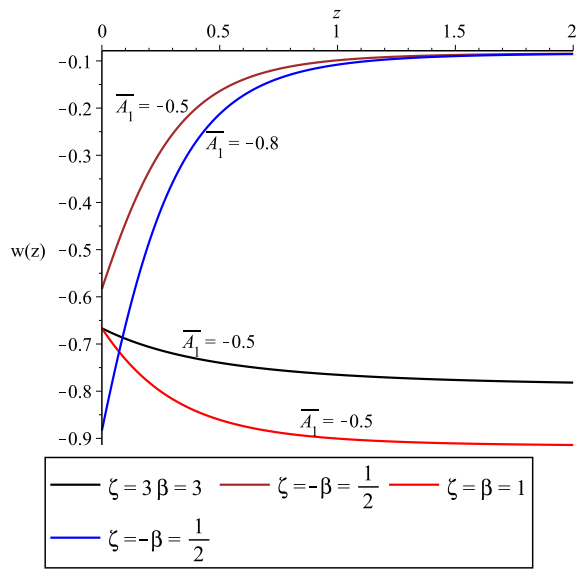}
\caption{EoS given by Eq. (\ref{pressub3}) as a function of the
redshift for some values of $\beta$, $\zeta$ and $\bar{A}_1$.}
\end{figure}

It is worth mentioning that, in the $z\rightarrow\infty$ limit,
$w(z)$ converges to $-\frac{\beta}{\zeta+3}$ while
$\frac{3(\zeta+3-\beta)}{2\zeta}\leq0$, and  converges to
$\bar{A_1}(\frac{\zeta+3-\beta}{\zeta+2-\beta})-\frac{\beta}{\zeta+3}$
if $\frac{3(\zeta+3-\beta)}{2\zeta}\geq0$. Moreover, as it is
clear from Fig~($3$), the $\zeta=3\beta=3$ and $\zeta=\beta=1$
cases, whenever $\bar{A}_1=-0.5$, may be used to unify the
dominated fluids in both the primary inflationary and current
accelerating phases into one model.

\begin{figure}[ht]
\centering
\includegraphics[scale=0.4]{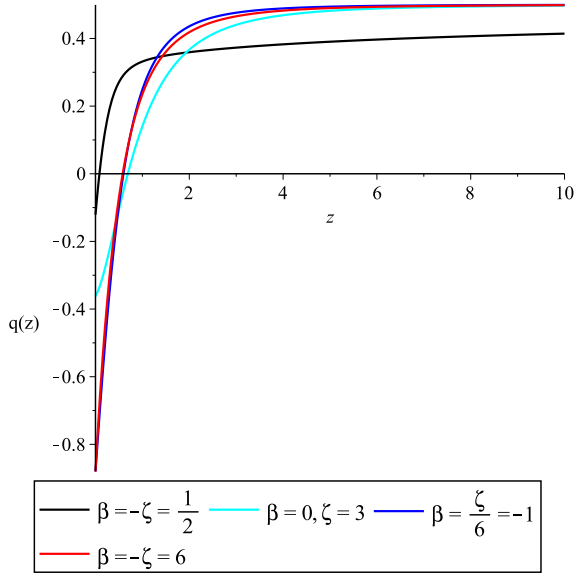}
\caption{ Evolution of the deceleration parameter, Eq.
(\ref{pressub5}), as a function of the redshift.}
\end{figure}
\noindent $q(z)$ has also been plotted in Fig.~($4$) for some
values of $\zeta$ and $\beta$, whenever $\Omega_{m0}=0.31$. All
the curves converge to $\frac{1}{2}$ at the $z\rightarrow\infty$
limit. Transition from the decelerating universe to an
accelerating universe happens around $z\simeq0.6$ for the
$\beta=\frac{\zeta}{6}=-1$ case. Besides, such transition is
happen around $z\sim0.62$ and $z\sim0.69$ for the $\beta=-\zeta=6$
and ($\beta=0, \zeta=3$) cases, respectively. Finally, it is
useful to mention that this transition also happens around
$z\sim0.082$ for the $\beta=-\zeta=\frac{1}{2}$ case.




\section{From the equation of state, Friedmann equation and thermodynamic pressure to energy}

As we have mentioned in the introduction, the study of the
cosmological features of fluids of $p=p(\rho)$ is important
\cite{prd0,capa,wang}. Therefore, we here consider a case in which
the $p(\rho)$ relation is given, such as the Polytropic and
Chaplygin models. One can use Eq.~(\ref{rhov}) to obtain $p(V)$.
In addition, by inserting the result into Eq.~(\ref{pres1}) and
using Eq.~(\ref{fried1}) one can obtain an expression for the
energy as a function of $\rho$ and $V$. As an example, if we
assume that $p=\omega\rho$ and combine Eqs.~(\ref{fried1})
and~(\ref{pres1}) with each other, we have that $E=-3\omega\rho V$
relation for energy, as we have previously obtained in the second
section, while $\omega$ is constant. Therefore, solutions with
$\omega\leq0$ leads to positive energy, a result which may be in
line with the current accelerating universe. However, it is in
direct conflict with the radiation and matter dominated eras as
well as the thermodynamic stability conditions \cite{rafael}. One
can check to see that the $\rho(a)$ and $a(t)$ relations are the
same as those of the standard cosmology. It is also useful to note
that the authors in \cite{rafael}, considered the total mass
definition for energy ($E=\rho V$) and they had shown that a dark
energy fluid satisfying $p=\omega\rho$, which has only one free
parameter including $\omega$, is unphysical. In addition,
following the above recipe for $p=K\rho^\gamma$ when $K$ and
$\gamma$ are constant. We have
$$E=\Big[\frac{3}{2\gamma-3}\Big]\Big(\frac{3}{32\pi}\Big)^{\gamma-1}K(\rho
V)^{3-2\gamma}$$ for the energy which, as a check, covers the
$p=\omega\rho$ case in the appropriate limit $\gamma=1$ and
$K=\omega$. The $E>0$ condition also leads to $\gamma>\frac{3}{2}$
for $K>0$ and $\gamma<\frac{3}{2}$ for $K<0$. Here, it is useful
to note that these solutions, which have two free parameters
including $K$ and $\gamma$, may satisfy the thermodynamic
stability conditions \cite{abri,plb1,plb2,ptp}.


Let us now consider the general case $p=K\rho^{\gamma}$ and insert
it into Eq.~(\ref{cont1}) to obtain
\begin{eqnarray}\label{den3}
\rho(a)=\rho_0\Big[a^{3(\gamma-1)}-K\Big]^{\frac{1}{1-\gamma}},
\end{eqnarray}
for density. By using Eq.~(\ref{rhov}), we can write that
\begin{eqnarray}\label{v1}
V(a)=V_0\Big[a^{3(\gamma-1)}-K\Big]^{\frac{3}{2(\gamma-1)}},
\end{eqnarray}
where
$V_0=2(\frac{32\pi}{3})^{-\frac{1}{2}}\rho_0^{-\frac{3}{2}}$. We
can see that for $\frac{3}{2}<\gamma$, density and volume are
decreased and increased, respectively. Now, let us study the
behavior of the deceleration factor $q=-1-\frac{\dot{H}}{H^2}$.
Since $1+z=\frac{1}{a}$, where $z$ is the redshift, by using
Eqs.~(\ref{fried1}),~(\ref{ray}) and~(\ref{den3}) we can write
that
\begin{eqnarray}\label{q1}
q=\frac{1}{2}[1+\frac{K'(1+z)^{3(\gamma-1)}}{1-K(1+z)^{3(\gamma-1)}}],
\end{eqnarray}
where $K'=3K\rho_0^{\gamma-1}$. For $\gamma>\frac{3}{2}$, while
$E>0$, we reach
\begin{eqnarray}\label{q10}
q\sim\frac{1}{2}\Big[1-\frac{K'}{K}\Big],
\end{eqnarray}
and
\begin{eqnarray}\label{q101}
q\sim\frac{1}{2}\Big[1+\frac{K'}{1-K}\Big],
\end{eqnarray}
for the $z\rightarrow\infty$ and $z\rightarrow0$ limits,
respectively. Bearing the deceleration parameter of radiation
matter era ($q=1$) in mind, by implying this condition to
Eq.~(\ref{q10}) we have
$\rho_0=(\frac{-1}{3})^{\frac{1}{\gamma-1}}$. In order to have an
estimation for $K$, we insert the $q=-0.6$ value for the
$z\rightarrow0$ limit \cite{roos}, leading to $K=\frac{11}{16}$,
which is compatible with the $E>0$ and $\gamma>\frac{3}{2}$
conditions, and thus $K'=-\frac{11}{16}$. Finally, we have
\begin{eqnarray}\label{q11}
q=\frac{1}{2}\Big[1-\frac{11(1+z)^{3(\gamma-1)}}{16-11(1+z)^{3(\gamma-1)}}\Big],
\end{eqnarray}
where $\frac{3}{2}<\gamma$. Recent observations indicate that
there is a transition from deceleration to acceleration phase at
$z\simeq0.6$ in the universe history \cite{obs1,obs2,obs3}. Using
this observational result in Eq.~(\ref{q11}) we have that
$\gamma\simeq0.77$ which is in contrast with $\frac{3}{2}<\gamma$
and thus the $E>0$ expectation. In addition, one can check to see
that the $q(z\rightarrow\infty)\rightarrow-1$ initial condition,
leads to $K<0$ for $\gamma>\frac{3}{2}$ which is unsatisfactory.
Moreover, by applying the
$q(z\rightarrow\infty)\rightarrow\frac{1}{2}$ initial condition
leads to two solutions. One of them yields $K=0$ which covers the
matter dominated era ($q=\frac{1}{2}$ and $\rho(a)=\rho_0a^{-3}$)
but with $E=0$.   The other leads to the unphysical solution
$\rho(a)=0$. Therefore, these type of solutions
($\frac{3}{2}<\gamma$) seem unsatisfactory.

Now, let us focus on the $\gamma<\frac{3}{2}$ case, where the
$E>0$ condition requires that $K$ should be a negative quantity.
For $z\rightarrow\infty$, we can write that $q\simeq\frac{1}{2}$,
which signals us to a matter dominated era \cite{roos}. In order
to continue our investigation, we confine ourselves to the
$\rho_0=1$ case yielding $K'=3K$. Additionally, since
$q(z\rightarrow0)\simeq-0.6$ \cite{roos} and
$q(z\simeq0.6)\simeq0$ \cite{obs1,obs2,obs3}, we have
$K=\frac{K'}{3}=-\frac{11}{4}$ and $\gamma\simeq-0.21$. Therefore,
a solution with $\rho_0=1$, $K=-\frac{11}{4}$ and
$\gamma\simeq-0.21$ can satisfy both the theoretical ($E>0$) and
observational ($q(z\rightarrow0)\simeq-0.6$ and
$q(z\simeq0.6)\simeq0$) conditions. Moreover, simple calculations
show that this model is free from any future singularity. Finally,
as it is shown in Fig~($4$), this fluid may be used to describe
the universe history from the matter dominated era to its current
phase.

\begin{figure}[ht]
\centering
\includegraphics[scale=0.4]{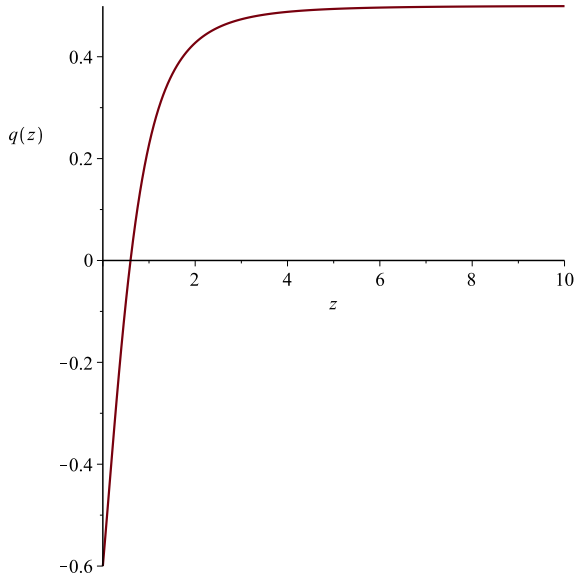}
\caption{Evolution of the $q(z)$ function.}
\end{figure}

Moreover, since $E\sim(\rho V)^{3-2\gamma}\sim
V^{\frac{3-2\gamma}{3}}$, by simple calculations we obtain
$E\sim\frac{1}{((1+z)^{3.63}+\frac{11}{4})^{1.41}}$, for the
energy evolution in this model. In order to have a vision about
the energy changes in this model, we have plotted
$E=E_0\frac{1}{((1+z)^{3.63}+\frac{11}{4})^{1.41}}$ for $E_0=1$,
in Fig~($5$).

\begin{figure}[ht]
\centering
\includegraphics[scale=0.4]{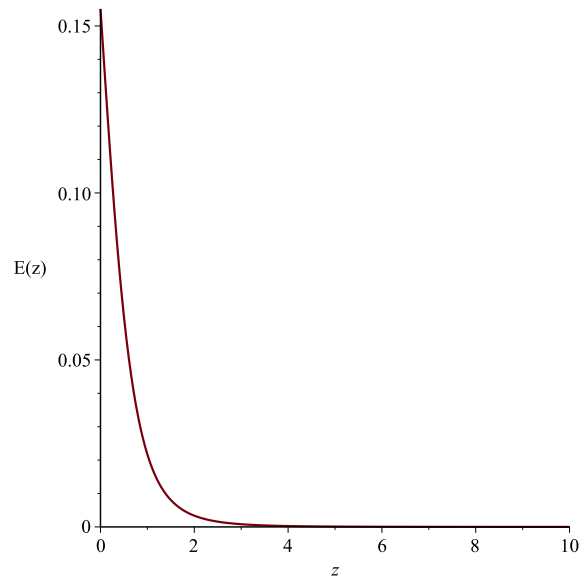}
\caption{Energy, $E(z)$, of the polytropic gas as a function of
the redshift.}
\end{figure}

However, since $E\rightarrow0$ in the $z\rightarrow\infty$ limit,
it seems that this solution is unsatisfactory for describing the
matter dominated era, where $q=\frac{1}{2}$. Therefore, this type
of solution may be used to describe a universe with
$q<\frac{1}{2}$.


\section{Observation constraints}

In this section we discuss the observational constraints on the
cosmological fluid defined in the last section, i.e., the density
energy given by Eq. (\ref{den3}). Let us assume that this fluid
can be interpreted with a model of unification in the dark sector,
i.e, cosmological scenario that unifies dark matter and dark
energy. Such proposal has been considered by many authors in the
literature
\cite{chap1,chap10,DE,DE0,DE1,DE2,DE3,DE4,DE5,DE6,DE7,DE8,DE9,DE10,DE11,DE12,DE13,DE14}.
Thus, with these considerations the Hubble expansion rate can be
written as

\begin{eqnarray}\label{H}
&& H(z) \\
&&\!\!\!\!\!= H_0 \sqrt{ \Omega_{b0}(1+z)^3 +
(1-\Omega_{b0})[(1+z)^{-3(\gamma-1)}-K]^{\frac{1}{1-\gamma}}}
\nonumber
\end{eqnarray}

To fit the free parameters $K$, $\lambda$, and $h$\footnote{Here
the parameter $h$ represent the reduced Hubble constant valued at
present moment, i.e. h = $H_0$/100.}, we can use the public code
CLASS \cite{class} in connection to the Monte Carlo code public
Monte Python \cite{monte}. We choose the Metropolis Hastings
algorithm as our sampling method. The data set utilized in our
analysis are: a) Type Ia supernovae (SNe Ia) from Union 2.1
compilation \cite{u21}, available at
http://supernova.lbl.gov/Union, that contains 580 SN Ia data in
the redshift range $0. 015 \leq z \leq 1.41$; b) Baryon acoustic
oscillations (BAO) data measurement from  the  Six  Degree  Field
Galaxy  Survey \cite{bao1}, the  Main  Galaxy  Sample  of  Data
Release 7  of  Sloan  Digital  Sky  Survey \cite{bao2}, the  LOWZ
and  CMASS  galaxy  samples  of  the Baryon  Oscillation
Spectroscopic  Survey \cite{bao3},  and the distribution of the
Lyman Forest in BOSS \cite{bao4}; c) Observational Hubble
parameter (OHD) data compiled by X. L Meng et al. \cite{hdata},
which comprising in 37 data points in the in the redshift range
$0.0708 \leq z \leq 2.36$. During the realization of statistical
analysis we consider $\Omega_{b0} = 0.05$ and $|K|$.
\\

\begin{figure}
  \includegraphics[width=3.0in, height=3.0in]{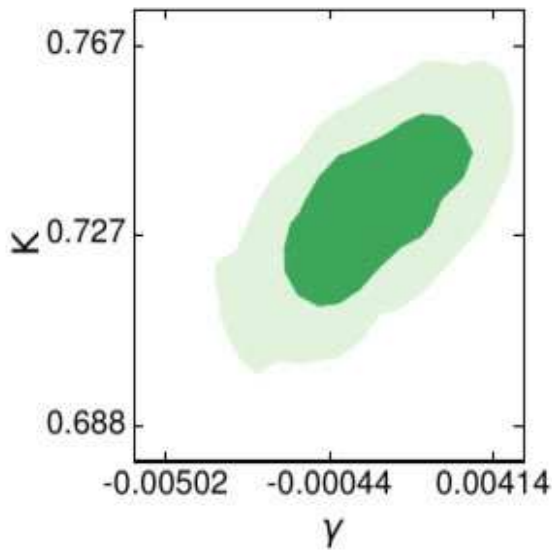}
  \caption{\label{results_ps1}  1$\sigma$ and 2$\sigma$ confidence level in the parametric space $\gamma - K$
  from the joint analysis SNe Ia + BAO + OHD.}
\end{figure}

\begin{figure}
  \includegraphics[width=1.5in, height=1.5in]{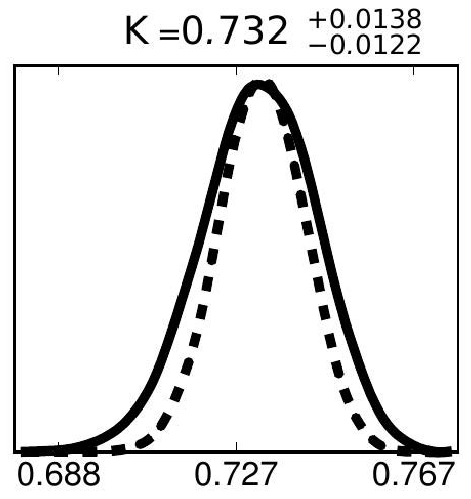}
  \includegraphics[width=1.5in, height=1.5in]{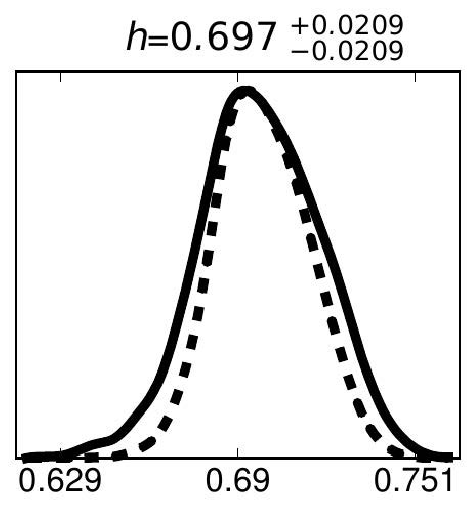}
  \includegraphics[width=1.5in, height=1.5in]{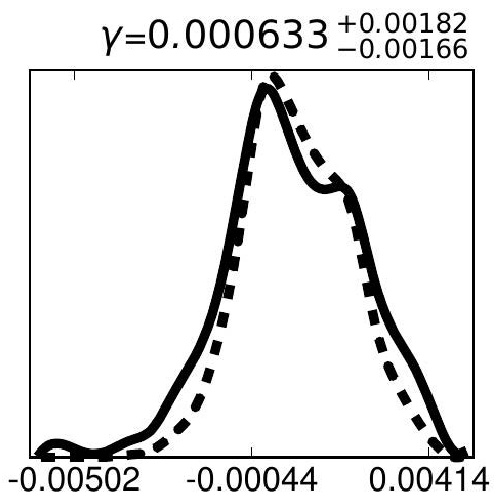}
 \caption{\label{results_ps2}
Marginalized one-dimensional posterior distribution for the
parameters $K$, $h$, and $\gamma$. The dashed line stands for the
average likelihood distribution.}
\end{figure}

Figure \ref{results_ps1} shows the confidence regions at 1$\sigma$
and 2$\sigma$ C. L in the plane $\gamma - K$ for the proposed
cosmological model. Figure \ref{results_ps2} shows the
marginalized one-dimensional posterior distribution for the
parameters of the models, with their corresponding 1$\sigma$
uncertainties. It is easy to see that for $\gamma = 0$ the Hubble
expansion rate defined in Eq. (\ref{H}) has a similar dynamic to
the $\Lambda$CDM model, if we assume that $1-\Omega_{b0} \simeq
1$. In this case the constant $K$ plays the role of a
``cosmological constant''. Our results summarized in Figures
\ref{results_ps1} and \ref{results_ps2} show that a polytropic gas
with EoS given by $p = K \rho^{\gamma}$ has a obvervacional fit
compatible with the $\Lambda$CDM model. Showing that a
cosmological unified scenario for dark matter and dark energy, and
which satisfies the thermodynamic conditions presented in the
present work, can fit the current astrophysicists data in the
presence of a single dark fluid.

\section{Summary and discussion}

In the standard cosmology, a perfect fluid with $p= w \rho$ and a
constant equation of state ($w$), satisfying the Friedmann and
continuity equations simultaneously, can be used to construct a
model for the universe expansion. Our analysis shows that, during
the cosmos evolution, it is impossible to preserve the Friedmann
and the thermodynamic pressure equations simultaneously, if one
takes into account the perfect fluid concept. In addition, we have
referred to the Komar mass and found that if we combine it with
the Friedmann first equation and insert the result into the
thermodynamic pressure equation, we construct an equation of state
which is similar to the modified polytropic and modified Chaplygin
models. Following that, by using the continuity equation, we can
write a relation for the density profile of cosmic fluid which may
be in agreement with the primary inflationary and current phases
of the universe expansion. In fact, the obtained density is free
from singularity and it may unify the nature of the dominated
fluid in both the primary inflationary and current accelerating
eras into one fluid. Moreover, we have considered a special
expression for energy that covers the previous definitions.
Combining this energy definition with the thermodynamic pressure
definition and the Friedmann second equation, we got the $p(V)$
relation. Bearing the Friedmann first equation in mind, we could
obtain the $p(\rho)$ relation, and showed that the obtained
solutions can be used to model the universe expansion, a result
which is in agreement with some previous attempts
\cite{od1,nojp,od2,od3,poly1,ref57,nK,poly2,poly20,poly200,chap1,chap10,chap2,chap3,van,jenk,jenk1,nucl}.
We also found out that some type of the obtained solutions may be
used to unify the dominated fluids in both the primary
inflationary and current accelerating phases into one model. In
addition, our study helps us to obtain the corresponding energy of
these models, in the cosmological setup, in a similar line with
the thermodynamic pressure definition. Therefore, our
investigation may also be considered as the thermodynamic
motivation for these models. Thereinafter, we have focused on the
situation in which the $p(\rho)$ relation is valid. By combining
the equation of state with the Friedmann first equation and
inserting the result into the pressure thermodynamic definition,
we could obtain a relation concerning the energy. Our study shows
that this recipe does not lead to a suitable result for the
perfect fluid ($p=w\rho$) case. In addition, we saw that the
$p=K\rho^{\gamma}$ case, in which $K$ is constant, leads to a
solution which may be free from any future singularity and also
may govern the universe expansion during the time that its
deceleration parameter meets the $q<\frac{1}{2}$ condition. We
have constrained the parameters $K$ and $\gamma$ with recent
observational data from SNe Ia + BAO + OHD. Based on our results,
solving the Friedmann and the thermodynamic pressure equations
simultaneously, one gets solutions for cosmic fluids that can be
used to model the dark energy candidates. Finally, it is
worthwhile mentioning that our investigation suggests that the
thermodynamic pressure equation, as a thermodynamic equation of
state, imposes serious constraints on the cosmic fluids (solutions
of the Friedmann equations), their energy and thus the energy
definitions. Therefore, this is still an unanswered question
dealing with what kind of fluids and energy definition one can
solve the Friedmann and thermodynamics equations simultaneously.
The latter being due to the fact that we do not exactly know the
true final forms of the laws governing the universe expansion and
energy definition in the gravitational theories. We also think one
may use the thermodynamic pressure definition to write a relation
for the energy of cosmic fluid in the cosmological setups.



\noindent


\end{document}